\documentclass[11pt]{article}
\usepackage[preprint]{acl}
\usepackage{times}
\usepackage{threeparttable}
\usepackage{enumerate}

\usepackage[T1]{fontenc} 
\usepackage{graphicx}
\usepackage{xspace}
\usepackage{url}
\usepackage{amssymb}
\usepackage{multirow}
\usepackage{xcolor}
\usepackage{colortbl}
\usepackage{tcolorbox}
\tcbuselibrary{breakable}  
\tcbuselibrary{skins}      
\usepackage{booktabs}

\usepackage{listings}
\definecolor{codeblue}{RGB}{78, 156, 205}
\definecolor{codegreen}{RGB}{96, 139, 78}
\definecolor{codepurple}{RGB}{197, 134, 192}
\definecolor{codeorange}{RGB}{206, 145, 120}
\definecolor{codegray}{RGB}{128, 128, 128}

\lstdefinelanguage{JavaScript}{
  morekeywords={break, case, catch, continue, debugger, default, delete, do, else, finally, for, function, if, in, instanceof, new, return, switch, this, throw, try, typeof, var, void, while, with, let, const, async, await, class, extends, import, export, unknown, string, is, typeof},
  morecomment=[l]{//},
  morecomment=[s]{/*}{*/},
  morestring=[b]',
  morestring=[b]",
  morestring=[b]/, 
  sensitive=true
}

\usepackage{array}
\usepackage{fontawesome5}
\usepackage{graphicx}
\usepackage{enumitem}
\usepackage{amsmath}
\usepackage{pifont}
\graphicspath{ {./images/} }

\newcommand{\vulbench}{\textsc{SastBench}\xspace}
\newcommand{\vulbenchVersion}{\vulbench-v0.1\xspace}
\newcommand{\xmark}{\textcolor{red!70!black}{\ding{55}}}
\renewcommand{\checkmark}{\textcolor{green!70!black}{\ding{51}}}
\newcommand{\halfcheckmark}{\textcolor{orange!70!black}{\ding{51}\kern-1.1ex\raisebox{.7ex}{\rotatebox[origin=c]{125}{\textbf{--}}}}}
\title{\vulbench: A Benchmark for Testing Agentic SAST Triage}
\author{
\textbf{Jake Feiglin \& Guy Dar} \\
{Rival Labs} \\
\texttt{\{jake, guydar\}@rival.security} \\[0.42em]
\includegraphics[height=0.34cm]{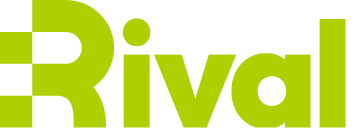} 
}
\date{}

\begin{document}

\maketitle
\begin{abstract}
SAST (Static Application Security Testing) tools are among the most widely used techniques in defensive cybersecurity, employed by commercial and non-commercial organizations to identify potential vulnerabilities in software. Despite their great utility, they generate numerous false positives, requiring costly {manual filtering} (aka \textit{triage}). While LLM-powered agents show promise for automating cybersecurity tasks, existing benchmarks fail to emulate real-world SAST finding distributions. We introduce \vulbench, a benchmark for evaluating SAST triage agents that combines real CVEs as true positives with filtered SAST tool findings as approximate false positives. \vulbench features an agent-agnostic design. We evaluate different agents on the benchmark and present a comparative analysis of their performance, provide a detailed analysis of the dataset, and discuss the implications for future development. 
\end{abstract}
\section{Introduction}
SAST ({Static Application Security Testing}) tools help prevent security risk in production systems by automatically scanning the source code for potential security vulnerabilities before they reach production. However, a critical and well-documented limitation of SAST tools is their propensity to generate a high volume of false positives -- alerts that flag benign code as vulnerable. This noise creates a significant burden for security analysts, who must manually triage each finding; i.e., to distinguish true vulnerabilities from false alarms. This process is slow, expensive, and prone to human error, often leading to alert fatigue where legitimate threats may be overlooked.

Recent advances in Large Language Models (LLMs) have demonstrated remarkable capabilities in code comprehension, reasoning, and analysis. The development of LLM-powered autonomous agents is a promising direction for automated SAST triage. By equipping these agents with tools to navigate code repositories, analyze context, and reason about code behavior, we can anticipate a future where they serve as intelligent first responders, accurately validating vulnerabilities, and dramatically reducing the manual overhead for humans.
Many cybersecurity datasets have emerged to test AI models for vulnerability detection and classification. They employ different approaches, but the data distribution is rarely designed to simulate the task of triaging SAST findings. Consequently, datasets miss important aspects of triage (see Section~\ref{subsec:existing_benchmarks}). This discrepancy leaves engineers guessing as to the performance of their auto-triage tools in the wild. 
\begin{table}[t]
\centering
\begin{threeparttable}
\tiny
\setlength{\tabcolsep}{2pt}
\hspace{-0.3em}\begin{tabular}{@{}lccccccc} 
\toprule
\textbf{Benchmark} & \textbf{Post-Cutoff} & \textbf{Agentic} & \textbf{Language} & \textbf{Scale \&} & \textbf{Hard Neg. /} & \textbf{Realistic FP} \\ 
 & \textbf{Data} & \textbf{Design} & \textbf{Diversity} & \textbf{Diversity} & \textbf{Paired Setup} & \textbf{Distribution}\\ 
\midrule
Juliet & \xmark & \xmark & \xmark & \halfcheckmark\tnote{1} & \checkmark & \xmark \\ 
CASTLE & \xmark & \xmark & \xmark & \xmark & \checkmark & \xmark \\ 
Devign & \xmark & \xmark & \xmark & \halfcheckmark\tnote{2} & \checkmark & \xmark \\ 
CVEFixes & \xmark & \xmark & \checkmark & \checkmark & \xmark & \xmark \\ 
PrimeVul & \xmark & \xmark & \xmark & \checkmark & \checkmark\tnote{3} & \xmark \\ 
DiverseVul & \xmark & \xmark & \xmark & \checkmark & \xmark & \xmark \\ 
ReposVul & \xmark & \xmark & \halfcheckmark\tnote{4} & \checkmark & \xmark & \xmark \\ 
CleanVul & \xmark & \xmark & \checkmark & \checkmark & \checkmark & \xmark \\ 
VulEval & \xmark & \xmark & \xmark & \checkmark & \xmark & \xmark \\ 
JitVul & \xmark & \checkmark & \xmark & \checkmark &\checkmark & \xmark \\ 
eyeballvul & \checkmark & \checkmark & \checkmark & \checkmark&  \xmark & \xmark \\ 
\midrule
\textbf{\vulbench} & \checkmark & \checkmark & \checkmark & \checkmark & \checkmark & \checkmark \\ 
\bottomrule
\end{tabular}
\caption{Comparison of \vulbench with a selection of representative existing benchmarks.}
\label{tab:benchmark_comparison}
\begin{tablenotes}
\item[1] Limited number of templates
\item[2] Limited number of repositories
\item[3] They provide two variants -- one of them is paired.
\item[4] Python, Java, C/C++; misses important languages such as PHP, Go, and Javascript/Typescript. 
\end{tablenotes}
\end{threeparttable}
\end{table}

\paragraph{Why SAST triage?} SAST tools are already a core cyberdefense strategy of companies. They effectively serve as a first filtering step, focusing the efforts of security analysts by obviating the need to scan the entire code. From a practical standpoint this helps us too, as automators, to reduce the agents' workload. Further, due to the prevalence of SAST triage in the industry, integration with existing pipelines is seamless. Unsurprisingly, many AI tools and products have emerged with the promise of vulnerability triage, \textbf{but without a suitable, agreed-upon dataset, progress cannot be truly tracked}, and tools cannot be compared.  From an evaluation point of view, the assessment of SAST triage is a more well-posed question than vulnerability detection; e.g., metrics may be misleading -- most benign code is easy to dismiss, inevitably leading to an overestimation of model capabilities, while SAST findings are harder to classify, forming a {hard} negative class. 

In this paper, we introduce \vulbench, a new benchmark designed specifically to evaluate the ability of LLM agents to classify SAST findings. \vulbench aims to minimize the simulation-reality gap prevalent in existing benchmarks. 
While it is easy to sample from the distribution of SAST findings $\mathcal{D}_\text{SAST}$ (simply running the tool on publicly available code), it is hard to sample at scale from the distribution of SAST false positives $\mathcal{D}^\text{FP}_\text{SAST}$ and SAST true positives $\mathcal{D}^\text{TP}_\text{SAST}$ separately, as this would require solving the problem of triage automatically. As a proxy, we propose the following setup. 

First, we use the \textit{Common Vulnerabilities and Exposures} (CVE) database as a source for true positives. This provides a reliable source of human curated, community verified vulnerabilities. Second, we take the findings of a simple rule-based SAST tool, filter them according to simple heuristics and use as false positives. This forms a class of findings that are \textit{mostly} real false positives. 
Despite the potential existence of true positives among these findings, SAST tools are known to have a large proportion of false positives, and tend to be good at identifying artificially injected vulnerabilities, but not true vulnerabilities \citep{sate_vi_report}. Moreover, we increase this effect further by (a) using a simple SAST tool configuration rather than deep semantic understanding, and (b) filtering to minimize avoidable blunders.


The testing environment, too, is designed carefully to emulate the use case of an auto-triage agent. The design is unopinionated and agnostic to the way agents are designed, following agentic benchmarks like SWE-Bench \citep{swe_bench} and Terminal-Bench \citep{tbench_2025}. Agents have full access to the codebase and full freedom to explore the environment, and are evaluated solely based on their prediction. Competitors submit their agent as an arbitrary ZIP with a \verb|Dockerfile| and any number of arbitrary files with the only requirement that it exposes a REST endpoint to run the agent. The agent is loaded into an isolated environment with a target repository at a static path, and it is given a list of potentially vulnerable code sites, required to return a binary verdict: true positive or false positive. This design allows \vulbench to serve as a neutral testing ground for comparing diverse agentic design choices, models, and architectures.

We test the difficulty of \vulbench by conducting a comprehensive evaluation of various agentic paradigms, tools, and state-of-the-art LLMs. We find that stronger models tend to perform better in terms of both precision and recall, and that detailed security-oriented prompts improve performance dramatically. We open-source our code and data to foster community engagement and transparency in the pursuit of automating application security.

\section{Background}
\label{sec:background}
\subsection{SAST Tools}

SAST tools are designed to identify vulnerabilities in large codebases. Traditionally, SAST tools look for simple patterns (such as regular expressions), but these patterns miss important contextual information in the code. For example, a pattern match can suggest positions where user input may lead to command execution. But to understand whether it is exploitable, it requires both semantic and contextual understanding. For example, in Python/Flask projects, 99.5\% of flagged command injections were found to be false positives \citep{ghost2025cast}. This gap results in significant time spent on SAST triage, where security analysts review SAST findings and manually distinguish false alarms from true vulnerabilities. This task is very taxing and thankless, as most artifacts turn out to be false positives, with some estimates indicating 8\%-30\% security-related true positives \citep{sate_v_report}, depending on programming language. In a recent report covering 2,166 flagged vulnerabilities, \textbf{SAST tools generated 91\% noise} \citep{ghost2025cast}.  

\subsection{Existing Benchmarks}
\label{subsec:existing_benchmarks}
When analyzing existing vulnerability classification benchmarks, we have found that they are not well-designed for commercial automatic {triage} evaluation. Indeed, almost all were not even designed for triage. They were created for other tasks like detection or classification. Triage can also be seen as a classification task, but where data is sampled from a harder, more adversarial distribution -- $\mathcal{D}_\text{SAST}$ or an approximation of it. Samples from $\mathcal{D}_\text{SAST}$ make classification harder by design, as they are selected by their confusing, apparent vulnerability to SAST tools. Table~\ref{tab:benchmark_comparison} analyzes a list of representative datasets across several relevant aspects. We observe that many datasets lack at least some (and often most) of the following:
\begin{itemize}[itemsep=0pt,topsep=3pt]
    \item \textbf{False Positive Distribution}: Most benchmarks do not consider the task of triage explicitly. Therefore, an exaggerated set of (mostly) non-vulnerable functions form the false positive class. A few datasets consider a \textit{paired} setup, where false positives are generated from fixed code, but this does not well represent the false positive distribution of SAST findings either, because the fix itself may be emphasized, giving hints to the agent.  
    \item \textbf{Scale and Scope}: This problem is not as ubiquitous as other problems, but still common. We find that some datasets are curated manually, leading to small datasets that are not diverse enough to draw conclusions from. They are often useful for small, concentrated investigations, where authors can inspect behaviors directly. Some other datasets are derived from a small set of repositories, also leading to limited generalizability.
    \item \textbf{Language Diversity}: Datasets often concentrate on 1-4 programming languages. This does not reflect the language distribution in the wild. Moreover, the languages usually chosen (e.g., C/C++, Java) are not representative of the language distribution in modern target systems and in vulnerable systems in particular. For example, most datasets don't contain PHP, though it is one of the languages with most vulnerabilities and most SAST findings (see Figure~\ref{fig:dataset_language}).
    \item \textbf{Agentic Benchmark}: Most benchmarks do not take into account the affordances and challenges of agentic workflows, often simply incompatible, ignoring problems like data leakage from the parametric knowledge of the model -- though there's a recent trend that is more in tune with the agentic setup.  
\end{itemize}
Dividing datasets into broad categories, we identify roughly four categories:
\begin{itemize}[itemsep=0pt, topsep=2pt]
    \item \textbf{Synthetic/Manual}: These datasets are characterized by human-curated examples or templates. They are harder to generate and often focus on a narrow subset of the distribution. 
    \item \textbf{Detection Tasks}: Detection tasks give the entire codebase, either all at once, or one snippet at a time, to the model. This can be seen as a classification task with a very large set of false positives, often containing many easily identifiable ones. Therefore, precision is less useful and recall is the main metric. 
    \item \textbf{Paired Setup}: To avoid using the entire dataset as false positives, papers such as PrimeVul \citep{prime_vul} and JitVul \citep{jit_vul} generate a balanced dataset. True positives come from CVEs and false positives come from their \textit{fixes}. While very useful, this approach has two potential problems. First, the patch might not solve the problem hermetically or even introduce new bugs \citep{repos_vul}. Second, the fixing changes may hint at the fix, potentially making the task less adversarial than the SAST task. Even if it weren't the case, it still represents a biased sample from $\mathcal{D}^\text{FP}_\text{SAST}$. False positives, which are more critical to keep in-distribution to get a realistic triage estimation, are biased to a subset of false positives.  
    
    \item \textbf{SAST-based Setup}: Very few papers, such as D2A \citep{d2a_dataset} and Draper \citep{draper} use SAST findings in their pipeline. However, Draper uses SAST tools as the ground truth, which is antithetical to our needs (triage). D2A labels all SAST findings that do not disappear after a CVE fix as false positives, which is a slightly aggressive heuristic. For example, it can miss vulnerabilities from other CVEs in the repository. 
\end{itemize}

\section{Methodology}
\label{sec:methodology}
In this Section, we present the methodology used in \vulbench. Specifically, to facilitate consistent discussion about the benchmark, we use version numbers to help us keep track of changes made to the curation methodology. 
\textbf{In this version, we present \vulbenchVersion}.



\subsection{Data Curation}
\label{subsec:dataset}

The integrity of \vulbench hinges on reaching a good quality, realistic dataset. Our curation process ensures data points emulate a real SAST triage problem as faithfully as possible without incurring prohibitive curation costs. 

\paragraph{True Positives (Vulnerabilities).} We mine Common Vulnerabilities and Exposures (CVEs) from the National Vulnerability Database (NVD) that reference commits on GitHub with known reported vulnerabilities, based on CVEFixes \citep{cve_fixes} methodology. 
Each CVE is associated with a CWE (Common Weakness Enumeration) category, derived from a taxonomy system used to divide CVEs into broad categories. 

Based upon the observations of \citet{cve_bench}, we expect performance to depend on the model's knowledge cutoff. To avoid contamination, we keep CVEs only if they were reported after a \textbf{knowledge cutoff} period. In the instantiation used in this paper, we set it to February 2025, which is after the knowledge cutoff of all the models tested herein. However, we consider this part configurable and use it with the version tag of the benchmark (i.e., the full tag is \vulbench-v\verb|<number>|@\verb|<start_date>|-\verb|<end_date>|). This is another advantage of our framework -- the automated nature of our benchmark enables continuous updates in line with the philosophy of SWE-Bench \citep{swe_bench} and LiveBench \citep{live_bench}.

\paragraph{False Positives (SAST Findings).} We execute a popular open-source SAST tool -- semgrep's \textbf{free edition} -- on the pre-fix versions of the repositories containing our curated CVEs. Findings from the SAST tool are marked as the negative class if they don't share a CWE ID with the true positive of this commit hash. For each finding, we extract relevant information: file paths, line numbers, and CWE ID. 
To ensure high-quality negatives that reflect actual triage workloads, we apply a filtering step. 
We remove all findings in the same \textit{function} as an identified vulnerability (including from other vulnerabilities in this repository).
    

\paragraph{Dataset Format.}
A single entry in the dataset consists of a single commit hash-CWE pair, where all findings that are associated with this CWE are concatenated into one entry. For each commit hash, we aggregate all SAST findings by CWE. 
Analogously, we collect the true positive's affected lines and assign them the CWE ID provided in the CVE description. Each entry is a list of key-value dictionaries. Keys are detailed in Table~\ref{tab:dataset_fields}.

\begin{table}[h]
\centering
\scriptsize
\begin{tabular}{lp{4cm}}
\hline
\textbf{Field} & \textbf{Description} \\
\hline
\texttt{repo\_name} & The name of the repository \\
\texttt{commit\_hash} & The commit hash \\
\texttt{function\_name} & The name of the function \\
\texttt{function\_start\_line} & The starting line number of the function \\
\texttt{function\_end\_line} & The ending line number of the function \\
\texttt{finding\_start\_line} & The starting line number of the finding \\
\texttt{finding\_end\_line} & The ending line number of the finding \\
\texttt{language} & The programming language \\
\texttt{source} & \texttt{cve} or \texttt{semgrep} -- target feature, not passed to the agent. \\
\hline
\end{tabular}
\caption{Description of dataset fields}
\label{tab:dataset_fields}
\end{table}

\begin{figure*}[t]
    \centering
    \hspace*{-1.01em}
    \includegraphics[width=1.05\textwidth]{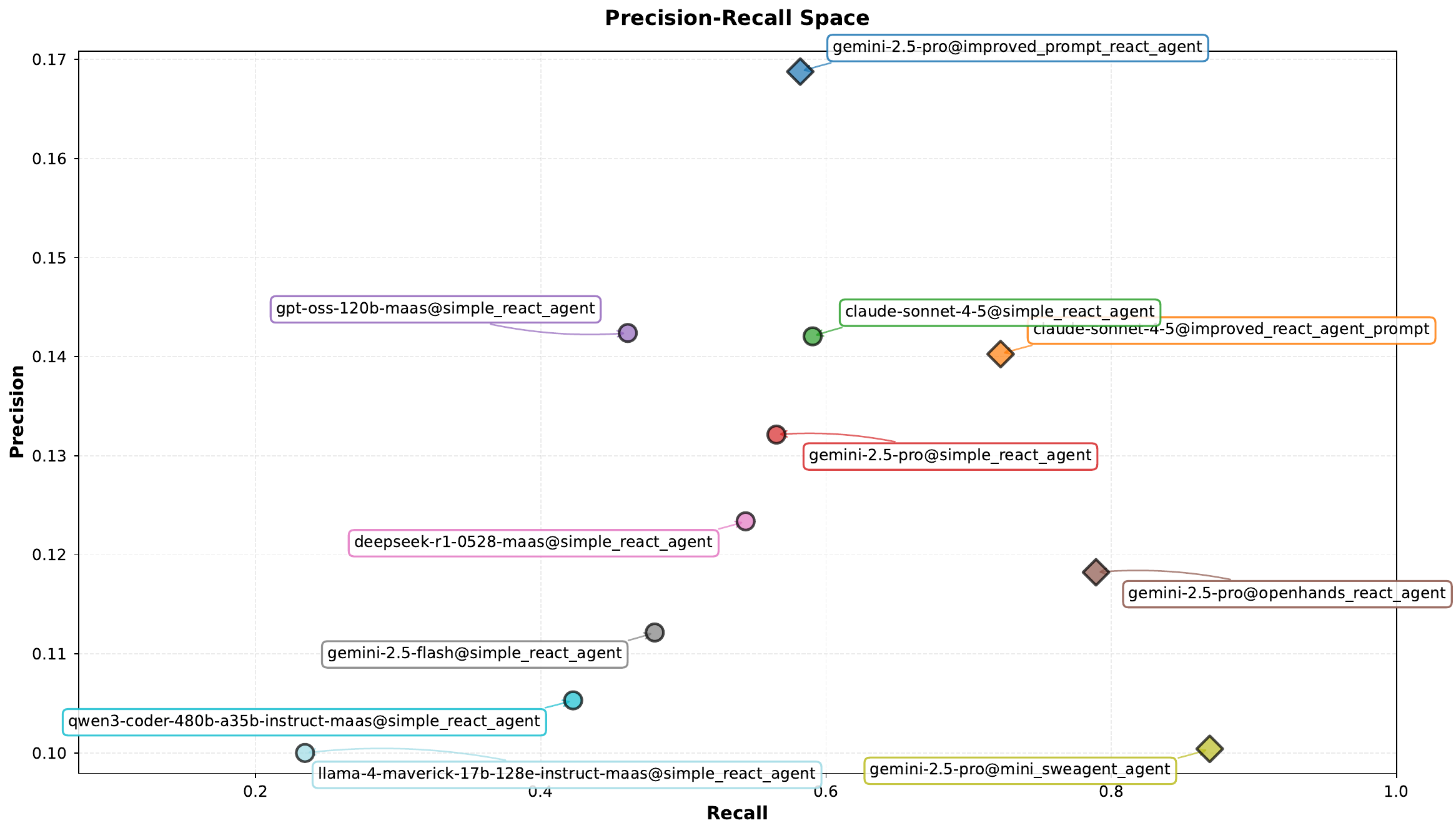}
    \caption{Precision-Recall space visualization. Points represent model-agent configurations, with position indicating trade-offs between false positive reduction (precision) and vulnerability detection (recall). Upper-right region represents ideal performance. Simple ReAct agents are indicated by circles, while other variants are presented as diamonds. To avoid clutter, we keep only the important instances.}
    \label{fig:precision_recall}
\end{figure*}
\subsection{Benchmark Design}
\label{subsec:design}

\paragraph{Task.} A Docker is spun up with the repository checked out to the provided commit hash, with one CWE group at a time, represented as a list of JSON objects as described above, excluding the \verb|source| feature, from which the target feature is derived. The agent's task is to execute its internal workflow to arrive at a binary decision. 
It must return a JSON object: \texttt{\{"verdict": "true\_positive" | "false\_positive"\}} and then the agent’s answer is compared with the ground truth. Performance is primarily evaluated based on \textit{Matthews' Correlation Coefficient} (MCC) due to the imbalanced nature of the task. We also report auxiliary metrics: precision, recall, F1, F2, and accuracy. 

\paragraph{Submission.} Competitors submit their agent as a single ZIP file containing a \texttt{Dockerfile} and all necessary source code. The code must implement a predefined API endpoint called \texttt{/analyze}. Our design imposes no constraints on the internal architecture of the agent, allowing for complete freedom in the choice of LLM, reasoning loops (e.g., ReAct), and tools (e.g., code browsers, compilers). 
For efficiency, we allow the user to define two regimes, instance-specific and commit-specific routines. The latter serves as a \textit{preprocessing stage}, run once per commit hash. This allows users to have commit-wide artifacts shared between instances readily available. The instance-level execution runs once for each record in the dataset and can use the created artifacts saved from the preprocessing stage.

\subsection{Dataset Composition}
Table~\ref{tab:data_statistics} summarizes important statistics of the dataset. 
In Appendix~\ref{app:additional}, several dataset statistics graphs are plotted. Figure~\ref{fig:dataset_language} presents the distribution across programming languages. The dataset spans languages commonly used in production systems, such as PHP, Javascript/Typescript, Python, and others. The data collection process (softly) enforces priors over programming languages related to their real-world frequencies.
Figure~\ref{fig:cwe_distribution} presents the CWE distribution, demonstrating coverage of common web vulnerabilities, memory safety issues, and logic flaws. 

\begin{table}[h]
    \centering
    \small
    \begin{tabular}{lr}
        \hline
        \textbf{Feature} & \textbf{Value} \\
        \hline
        Total Samples & 2737 \\
        True Positives & 299 \\
        False Positives & 2438 \\
        Imbalance Ratio & 8.15:1 \\
        Languages & 38 \\
        Unique CWEs & 139 \\
        \hline
    \end{tabular}
    \caption{Summary of dataset statistics}
    \label{tab:data_statistics}
\end{table}

\section{Experiments}
\label{sec:experiments}
\subsection{Experimental Setup}
\begin{table*}[h]
\centering
\scriptsize
\begin{tabular}{l l r r r r r r}
\toprule
\textbf{Model} & \textbf{Architecture} & \textbf{Acc.} & \textbf{Prec.} & \textbf{Recall} & $\textbf{F}_1$ & $\textbf{F}_2$ & \textbf{MCC} \\
\midrule
Gemini 2.5 Pro & Improved ReAct & 0.641 & \textbf{0.169} & 0.582 & \textbf{0.262} & \textbf{0.197} & \textbf{0.148} \\
Claude Sonnet 4.5 & Improved ReAct & 0.481 & 0.140 & {0.722} & 0.235 & 0.167 & 0.110 \\
Claude Sonnet 4.5 & Simple ReAct & 0.563 & 0.142 & 0.591 & 0.229 & 0.167 & 0.096 \\
Gemini 2.5 Pro & Mini SWE-Agent & 0.327 & 0.100 & \textbf{0.869} & 0.180 & 0.122 & 0.092 \\
Gemini 2.5 Pro & Simple ReAct & 0.567 & 0.140 & 0.565 & 0.224 & 0.165 & 0.084 \\
GPT OSS 120B\textsuperscript{*} & Simple ReAct & 0.642 & 0.142 & 0.461 & 0.218 & 0.165 & 0.083 \\
Gemini 2.5 Pro & No Tools & 0.450 & 0.128 & 0.692 & 0.216 & 0.153 & 0.072 \\
Gemini 2.5 Pro & OpenHands & 0.334 & 0.118 & 0.789 & 0.206 & 0.142 & 0.047 \\
DeepSeek R1 & Simple ReAct & 0.523 & 0.123 & 0.544 & 0.201 & 0.146 & 0.042 \\
Gemini 2.5 Flash & Simple ReAct & 0.521 & 0.112 & 0.480 & 0.182 & 0.132 & 0.007 \\
Qwen3 Coder 480B & Simple ReAct & 0.536 & 0.105 & 0.423 & 0.169 & 0.124 & -0.011 \\
Llama 4 Maverick 17B & Simple ReAct & \textbf{0.679} & 0.100 & 0.235 & 0.140 & 0.113 & -0.020 \\
\bottomrule
\end{tabular}
\caption{Model performance summary. Results are sorted by MCC (our primary metric). Bold values indicate best performance in each column.}
\end{table*}

\paragraph{Tools.} To test the difficulty of \vulbench, we have designed a series of preliminary tests evaluating different agentic paradigms and language models. Unless otherwise stated, the agents use the following tools:
\begin{itemize}[itemsep=0pt,topsep=4pt]
\item \textbf{Read File}: Reads a file
\item \textbf{List Dir}: Lists files in directory
\item \textbf{Search Symbol}: Searches a symbol in the codebase
\item \textbf{Security Patterns Tool}: A toy lookup table used to provide agents with a short sentence about specific CWEs. Mostly served as a distractor.
\end{itemize}

\paragraph{Agents.} We evaluate multiple configurations. 
We compare the workflows detailed below:
\begin{itemize}[itemsep=0pt,topsep=4pt]
    \item \textbf{No Tools Baseline}: We provide the LLM with a concatenated list of the relevant lines of code, with no access to tools, and use Chain-of-Thought (CoT; \citeauthor{wei2023chainofthought}, \citeyear{wei2023chainofthought}).
    \item \textbf{Simple ReAct Agent}: We use a ReAct loop \citep{react} with no optimizations.
    \item \textbf{Improved Prompt Agent}: A prompt designed with domain expertise for the ReAct agent (a researcher with security knowledge aided by an LLM). 
    \item \textbf{Generalist Agents}: We use generalist agents, OpenHands \citep{wang2025openhands} and mini-SWE-agent \citep{swe_agent} designed for general software developer tasks. Generalist agents use their own specialized set of tools instead of ours. 
\end{itemize}
Prompts are provided in Appendix~\ref{app:prompts} and agents are implemented with DSPy \citep{dspy}.
\paragraph{Models.} We compare different LLM backends including Llama Maverick 17B \citep{llama_maverick}, Gemini 2.5 Flash and Pro \citep{gemini25}, DeepSeek-R1 \citep{deepseek_r1}, Qwen3 Coder 480B \citep{qwen3, qwen25coder}, GPT-OSS 120B \citep{gpt_oss} and Claude Sonnet 4.5 \citep{anthropic2025sonnet45}, all capable of code understanding and reasoning. 

\paragraph{Costs.} Due to the size of the dataset, as well as the complexity of solving a single task, we limited the number of runs per model-architecture pair to one. Moreover, because our specific infrastructure relied on Google, we limited the non-simple experiments to mostly employ Gemini.

\begin{figure*}
    \centering
    \includegraphics[width=0.68\linewidth]{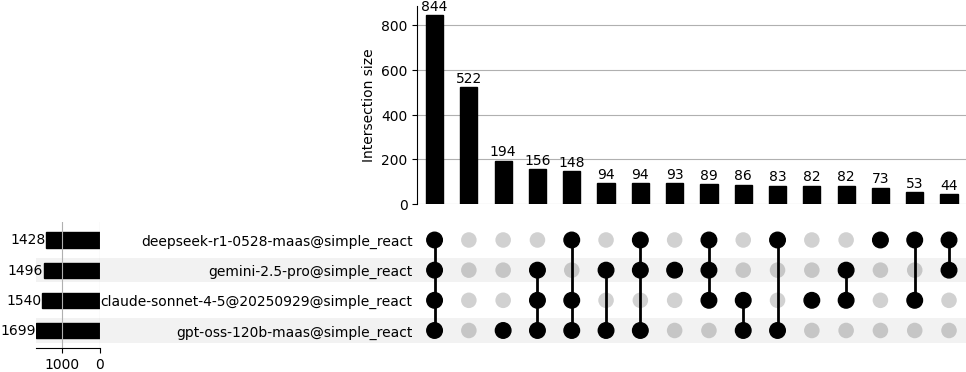}
    \caption{Upset plot for the Simple ReAct agent with different LLMs}
    \label{fig:upset_react}
\end{figure*}
\paragraph{Evaluation Metrics.}
We measure performance using standard classification metrics derived from the confusion matrix, which consists of four fundamental components: True Positives (TP), True Negatives (TN), False Positives (FP), and False Negatives (FN).
\textbf{Precision} is defined as $P = \mathrm{TP}/(\mathrm{TP}+\mathrm{FP})$ and measures the correctness of positive predictions. \textbf{Recall} is calculated as $R = \mathrm{TP}/(\mathrm{TP}+\mathrm{FN})$ and captures the model's sensitivity in detecting true positives. 

The \textbf{F\textsubscript{1} score} represents the harmonic mean of precision and recall, providing a balanced measure of both metrics, computed as $F_1 = 2PR/(P+R)$. For security-critical applications, it is also important to report metrics that weigh recall more heavily than precision. We use \textbf{F\textsubscript{2} score}, computed as $F_2 = 5PR/(4P+R)$. This reflects the reality that in vulnerability triage, missing a true vulnerability (false negative) typically incurs greater cost than raising a false alarm (false positive).

Finally, our primary evaluation metric is \textit{Matthews' Correlation Coefficient (MCC)}, which provides a balanced assessment across all four confusion matrix categories and remains robust under class imbalance:

{\footnotesize
\setlength{\abovedisplayskip}{5pt}
 \setlength{\belowdisplayskip}{10pt}
{
\vspace{-0.4em}
\begin{equation*}
\text{MCC} = \frac{\mathrm{TP}\cdot\mathrm{TN} - \mathrm{FP}\cdot\mathrm{FN}}
{\sqrt{(\mathrm{TP}+\mathrm{FP}) (\mathrm{TP}+\mathrm{FN}) (\mathrm{TN}+\mathrm{FP})  (\mathrm{TN}+\mathrm{FN})}}
\end{equation*}
\vspace{-0.8em}
}
}

MCC scores range from $-1$ (indicating total disagreement between predictions and ground truth) through $0$ (equivalent to random guessing) to $+1$ (representing perfect prediction). Unlike simple accuracy and even the $F_\beta$ scores (F1, F2), MCC remains a reliable indicator even in scenarios where true negatives significantly outnumber positive cases, making it particularly well-suited for vulnerability detection tasks where security issues are the outlier rather than default.

All experiments follow standardized evaluation protocols with identical inputs across models, ensuring that observed performance differences can be attributed solely to variations in model architecture rather than experimental conditions.
\subsection{Results}
Table~\ref{tab:benchmark_comparison} shows the results of the different models and architectures on the benchmark. 
Figure~\ref{fig:precision_recall} visualizes model behavior in terms of precision vs recall, capturing the trade-off between detection rate and false positive control. Each point represents a complete evaluation run, with annotated labels indicating specific configurations. Models near the upper-right corner excel at both detecting vulnerabilities and avoiding false positives, ideal for production deployment. Models with high recall but lower precision may suit high-security contexts where missing vulnerabilities is unacceptable, while high-precision models with moderate recall may be appropriate for resource-constrained environments where analyst time is limited.
Error patterns and classification behavior reveal that some configurations prioritize recall (fewer false negatives but more false positives), suitable for high-security contexts, while others balance precision-recall trade-offs, appropriate for resource-constrained environments.

\subsection{Analysis}
\paragraph{ReAct versus No-Tool Baseline.} Surprisingly, we find that Gemini without tool calls is almost on par with the simple agentic solution in terms of aggregate metrics (MCC, F1, F2) and exceeds in terms of recall (Table~\ref{tab:benchmark_comparison}), despite only relying on the immediate code locations and nothing else.  

\paragraph{Consistency Between Models.}
In Figure~\ref{fig:upset_react}, we show the correlation between the success and failure modes of all models in the Simple ReAct workflow design.  The Figure shows the agreement between models. It is clear from the plot that models tend to be all correct or all wrong together, indicating that they return similar verdicts. It suggests that they rely on similar reasoning and thinking patterns. Manually inspecting a handful of all-wrong results shows they are indeed wrong, and not mislabeled false positives.

\paragraph{Metrics are Correlated.} Figure~\ref{fig:precision_recall} shows that despite the inherent tension between precision and recall, they surprisingly tend to \textit{improve together} across Simple ReAct agents. Improvement is generally consistent with the trend one would predict a priori -- based on the ``strength'' of the models. In other words, stronger models tend to \textbf{pareto dominate} weaker models. GPT-OSS breaks away from this pattern, but interestingly, it is also the only model we had to run with a significant number of re-runs due to its struggles with DSPy.\footnote{GPT OSS runs often failed due to DSPy errors, {so we had to run many times} on ``bad'' inputs until a good run was completed. While these errors revolved around shallow formatting problems and not logical flow, it can still constitute implicit rejection sampling with potential bias.} 

It is also worth noting that aggregate metrics are also correlated. While accuracy, as expected, is useless and unrelated to the others -- F1, F2 and MCC produce almost the same model rankings. MCC still has the advantage of being more theoretically justified, as well as giving a clear way to compare to a random baseline (score zero).     

\paragraph{Improved Prompts Lead to Better Results.} We see that both Claude and Gemini with improved prompts are much better than their Simple ReAct counterparts. Curiously, Gemini improves in precision, while Claude improves in terms of recall. Generalist agents are not consistently better (at least with our minimal adjustments). 

In Appendix~\ref{app:example}, we provide an example walkthrough of two Claude agents on a true positive example -- the simple ReAct and improved prompt agent. The improved agent is able to identify that input validation has insufficient coverage.

\section{Related Work}
\label{sec:related_work}
\paragraph{Cybersecurity Benchmarks.} 
Early synthetic or hand-crafted benchmarks \citep{devign, sven, juliet_test_suite,castle} traded realism for inspectability. Most modern datasets however are constructed by mining vulnerability-fixing commits linked to public CVEs \citep{cve_fixes}, a process that scales well but introduces substantial label noise by conflating security-relevant and incidental code changes. 
Recent work has focused on exposing and mitigating the consequences of noisy labeling. PrimeVul \citep{prime_vul} demonstrates that evaluation practices often overestimate model performance and suggested mitigation strategies. CleanVul \citep{clean_vul} uses LLM-based analysis with heuristic filtering to clean CVE-mined datasets without full manual verification.

Beyond labeling quality, newer benchmarks emphasize broader context and realistic evaluation settings. ReposVul \citep{repos_vul} provides repository-level context across multiple languages, addressing tangled patches and inter-procedural dependencies. 
VulEval \citep{vul_eval} targets inter- and intra-procedural reasoning, while VulBench \citep{vul_bench} aggregates CTF datasets for quantitative evaluation of LLM-based vulnerability detection.

\paragraph{Agentic Benchmarks.} The evaluation of agentic AI systems has motivated development of specialized benchmarks that assess agentic task completion in realistic environments. 
SWE-Bench~\cite{swe_bench} evaluates language models on resolving real-world GitHub issues from open-source repositories, requiring agents to autonomously generate patches that pass existing test suites; subsequent refinements include SWE-Bench Verified~\cite{swebench_verified}, which addresses task quality through human validation, and SWE-Bench Pro~\cite{swebench_pro}, which increases difficulty through long-horizon complicated tasks.
Terminal-Bench~\cite{tbench_2025} focuses on command-line interface proficiency through tasks spanning code compilation, model training, and system debugging within containerized environments, while AssistantBench~\cite{assistant_bench} and WebArena~\cite{web_arena} evaluate agents on web tasks.
AgentBench~\cite{agent_bench} assesses agentic capabilities across diverse environments including operating systems, databases, and interactive platforms.
These benchmarks collectively reveal that while frontier models demonstrate strong performance on isolated tasks, substantial gaps persist in handling complex, multi-step workflows requiring sustained reasoning, tool use, and domain expertise.
In the cybersecurity domain, CVE-Bench~\cite{cve_bench} provides a real-world benchmark based on CVEs, where agents attempt to exploit vulnerabilities in environments that mimic production conditions, revealing that frontier models achieve limited success rates on genuine security exploits. 

\section{Discussion}
\label{sec:discussion}

Despite its limitations, we believe this benchmark has strong practical value for identifying \textit{generalizable} agentic workflows that align with real-world security triage. In practice, the task this benchmark emulates is not the discovery of novel vulnerabilities in isolation, but the identification of true security threats within large volumes of noisy SAST findings. 

We view the benchmark as a means toward this goal rather than an end in itself.
Importantly, maximizing benchmark performance is not necessarily difficult nor intrinsically meaningful. A trivial strategy could simply rerun the same SAST tool on the repository and check whether its findings coincide with the marked lines. However, such shortcut-based solutions do not reflect the intended use case. Our design philosophy emphasizes solutions that rely on justifiable reasoning trajectories rather than dataset-specific artifacts. We expect that models solving the task in this way will generalize more reliably to real-world security workflows.

To probe the presence of exploitable shortcuts, Appendix~\ref{app:shortcuts} evaluates the ability of simple, non-agentic classifiers to distinguish whether a sample originates from a CVE or from a SAST tool. These models are allowed to exploit shallow, learnable differences between the two data distributions. We find that such classifiers achieve only limited success, suggesting that non-contextual signals alone are insufficient to reliably separate the classes. This supports the claim that strong benchmark performance is unlikely to arise purely from shortcut learning.

\section{Conclusion}
In this paper, we present \vulbench, a scalable agentic benchmark for SAST triage. It is a first step toward democratizing the evaluation of auto-triage agents, and an alternative to  closed-source self-reports, which are hard to validate. We have analyzed different models and architectures on the benchmark and have demonstrated that stronger models and better prompts lead to better performance. We believe \vulbench is important for evaluation, but recognize its limitations. We recommend using our benchmark alongside other benchmarks (possibly with complementary problems) for a more complete picture.

\section{Limitations}
\label{sec:limitations}

To enable the automated construction of a scalable and practically useful dataset, we necessarily rely on heuristics. In particular, SAST findings are used as the negative class, reflecting the reality of industrial triage workflows, which are overwhelmingly dominated by SAST-generated noise. While it is theoretically possible that some SAST findings correspond to real but unreported vulnerabilities, several factors mitigate this risk.

First, SAST tools -- especially lightweight, pattern-based analyzers such as the one used in this work -- are well known to produce large numbers of false positives \citep{sate_v_report, sate_vi_report, ghost2025cast}. Moreover, even when SAST tools do identify true vulnerabilities, these are often of lower severity than those captured by CVEs. Second, the filtering heuristics applied during dataset construction further reduce the likelihood of contamination of the negative class with genuine vulnerabilities.

Empirically, we observe that stronger models tend to improve precision in addition to recall, despite the presence of noisy labels. This suggests that any noise affecting the precision metric is relatively small and does not overwhelm genuine performance differences between models. Nonetheless, the reliance on heuristic labeling remains an inherent limitation of the benchmark and should be considered when interpreting results.

\bibliography{references}

\begin{thebibliography}{37}
\providecommand{\natexlab}[1]{#1}

\bibitem[{Anthropic(2025)}]{anthropic2025sonnet45}
Anthropic. 2025.
\newblock \href {https://www.anthropic.com/news/claude-sonnet-4-5} {Introducing claude sonnet 4.5}.
\newblock Accessed: 2025-12-23.

\bibitem[{Bhandari et~al.(2021)Bhandari, Naseer, and Moonen}]{cve_fixes}
Guru Bhandari, Amara Naseer, and Leon Moonen. 2021.
\newblock \href {https://doi.org/10.1145/3475960.3475985} {Cvefixes: automated collection of vulnerabilities and their fixes from open-source software}.
\newblock In \emph{Proceedings of the 17th International Conference on Predictive Models and Data Analytics in Software Engineering}, PROMISE ’21, page 30–39. ACM.

\bibitem[{Boland and Black(2012)}]{juliet_test_suite}
Tim Boland and Paul~E. Black. 2012.
\newblock \href {https://doi.org/10.1109/MC.2012.345} {Juliet 1.1 c/c++ and java test suite}.
\newblock \emph{Computer}, 45(10):88--90.

\bibitem[{Comanici et~al.(2025)Comanici, Bieber, Schaekermann, Pasupat, Sachdeva, Dhillon, Blistein, Ram, Zhang, Rosen, Marris, Petulla, Gaffney, Aharoni, Lintz, Pais, Jacobsson, Szpektor, Jiang, Haridasan, Omran, Saunshi, Bahri, Mishra, Chu et~al.}]{gemini25}
Gheorghe Comanici, Eric Bieber, Mike Schaekermann, Ice Pasupat, Noveen Sachdeva, Inderjit Dhillon, Marcel Blistein, Ori Ram, Dan Zhang, Evan Rosen, Luke Marris, Sam Petulla, Colin Gaffney, Asaf Aharoni, Nathan Lintz, Tiago~Cardal Pais, Henrik Jacobsson, Idan Szpektor, Nan-Jiang Jiang, and 7 others. 2025.
\newblock \href {https://arxiv.org/abs/2507.06261} {Gemini 2.5: Pushing the frontier with advanced reasoning, multimodality, long context, and next generation agentic capabilities}.
\newblock \emph{Preprint}, arXiv:2507.06261.

\bibitem[{DeepSeek-AI et~al.(2025)DeepSeek-AI, Guo, Yang, Zhang, Song, Zhang, Xu, Zhu, Ma, Wang, Bi, Zhang, Yu, Wu, Wu, Gou, Shao, Li, Gao, Liu, Xue, Wang, Wu, Feng, Lu, Zhao, Deng, Zhang, Ruan, Dai, Chen, Ji, Li, Lin, Dai, Luo, Hao, Chen, Li, Zhang, Bao, Xu, Wang, Ding, Xin, Gao, Qu, Li, Guo, Li, Wang, Chen, Yuan, Qiu, Li, Cai, Ni, Liang, Chen, Dong, Hu, Gao, Guan, Huang, Yu, Wang, Zhang, Zhao, Wang, Zhang, Xu, Xia, Zhang, Zhang, Tang, Li, Wang, Li, Tian, Huang, Zhang, Wang, Chen, Du, Ge, Zhang, Pan, Wang, Chen, Jin, Chen, Lu, Zhou, Chen, Ye, Wang, Yu, Zhou, Pan, Li, Zhou, Wu, Ye, Yun, Pei, Sun, Wang, Zeng, Zhao, Liu, Liang, Gao, Yu, Zhang, Xiao, An, Liu, Wang, Chen, Nie, Cheng, Liu, Xie, Liu, Yang, Li, Su, Lin, Li, Jin, Shen, Chen, Sun, Wang, Song, Zhou, Wang, Shan, Li, Wang, Wei, Zhang, Xu, Li, Zhao, Sun, Wang, Yu, Zhang, Shi, Xiong, He, Piao, Wang, Tan, Ma, Liu, Guo, Ou, Wang, Gong, Zou, He, Xiong, Luo, You, Liu, Zhou, Zhu, Xu, Huang, Li, Zheng, Zhu, Ma, Tang, Zha, Yan, Ren, Ren, Sha, Fu, Xu, Xie, Zhang,
  Hao, Ma, Yan, Wu, Gu, Zhu, Liu, Li, Xie, Song, Pan, Huang, Xu, Zhang, and Zhang}]{deepseek_r1}
DeepSeek-AI, Daya Guo, Dejian Yang, Haowei Zhang, Junxiao Song, Ruoyu Zhang, Runxin Xu, Qihao Zhu, Shirong Ma, Peiyi Wang, Xiao Bi, Xiaokang Zhang, Xingkai Yu, Yu~Wu, Z.~F. Wu, Zhibin Gou, Zhihong Shao, Zhuoshu Li, Ziyi Gao, and 181 others. 2025.
\newblock \href {https://arxiv.org/abs/2501.12948} {Deepseek-r1: Incentivizing reasoning capability in llms via reinforcement learning}.
\newblock \emph{Preprint}, arXiv:2501.12948.

\bibitem[{Delaitre et~al.(2023)Delaitre, Black, Cupif, Haben, Alex-Kevin, Okun, Prono, and Delaitre}]{sate_vi_report}
Aurelien Delaitre, Paul~E. Black, Damien Cupif, Guillaume Haben, Loembe Alex-Kevin, Vadim Okun, Yann Prono, and Aurelien Delaitre. 2023.
\newblock \href {https://doi.org/10.6028/NIST.SP.500-341} {Sate vi report: Bug injection and collection}.

\bibitem[{Delaitre et~al.(2018)Delaitre, Stivalet, Black, Okun, Cohen, and Ribeiro}]{sate_v_report}
Aurelien Delaitre, Bertrand Stivalet, Paul Black, Vadim Okun, Terry Cohen, and Athos Ribeiro. 2018.
\newblock \href {https://doi.org/10.6028/NIST.SP.500-326} {Sate v report: Ten years of static analysis tool expositions}.

\bibitem[{Deng et~al.(2025)Deng, Da, Pan, He, Ide, Garg, Lauffer, Park, Pasari, Rane, Sampath, Krishnan, Kundurthy, Hendryx, Wang, Bharadwaj, Holm, Aluri, Zhang, Jacobson, Liu, and Kenstler}]{swebench_pro}
Xiang Deng, Jeff Da, Edwin Pan, Yannis~Yiming He, Charles Ide, Kanak Garg, Niklas Lauffer, Andrew Park, Nitin Pasari, Chetan Rane, Karmini Sampath, Maya Krishnan, Srivatsa Kundurthy, Sean Hendryx, Zifan Wang, Vijay Bharadwaj, Jeff Holm, Raja Aluri, Chen Bo~Calvin Zhang, and 3 others. 2025.
\newblock \href {https://arxiv.org/abs/2509.16941} {Swe-bench pro: Can ai agents solve long-horizon software engineering tasks?}
\newblock \emph{Preprint}, arXiv:2509.16941.

\bibitem[{Ding et~al.(2024)Ding, Fu, Ibrahim, Sitawarin, Chen, Alomair, Wagner, Ray, and Chen}]{prime_vul}
Yangruibo Ding, Yanjun Fu, Omniyyah Ibrahim, Chawin Sitawarin, Xinyun Chen, Basel Alomair, David Wagner, Baishakhi Ray, and Yizheng Chen. 2024.
\newblock \href {https://arxiv.org/abs/2403.18624} {Vulnerability detection with code language models: How far are we?}
\newblock \emph{Preprint}, arXiv:2403.18624.

\bibitem[{Dubniczky et~al.(2025)Dubniczky, Horvát, Bisztray, Ferrag, Cordeiro, and Tihanyi}]{castle}
Richard~A. Dubniczky, Krisztofer~Zoltán Horvát, Tamás Bisztray, Mohamed~Amine Ferrag, Lucas~C. Cordeiro, and Norbert Tihanyi. 2025.
\newblock \href {https://arxiv.org/abs/2503.09433} {Castle: Benchmarking dataset for static code analyzers and llms towards cwe detection}.
\newblock \emph{Preprint}, arXiv:2503.09433.

\bibitem[{Gao et~al.(2023)Gao, Wang, Zhou, Zhu, and Zhang}]{vul_bench}
Zeyu Gao, Hao Wang, Yuchen Zhou, Wenyu Zhu, and Chao Zhang. 2023.
\newblock How far have we gone in vulnerability detection using large language models.
\newblock \emph{arXiv preprint arXiv:2311.12420}.

\bibitem[{{Ghost Security}(2025)}]{ghost2025cast}
{Ghost Security}. 2025.
\newblock \href {https://reports.ghostsecurity.com/cast.pdf} {Exorcising the sast demons: Contextual application security testing (cast)}.
\newblock Technical report, Ghost Security.
\newblock CAST (Contextual Application Security Testing) research report.

\bibitem[{He and Vechev(2023)}]{sven}
Jingxuan He and Martin Vechev. 2023.
\newblock \href {https://doi.org/10.1145/3576915.3623175} {Large language models for code: Security hardening and adversarial testing}.
\newblock In \emph{Proceedings of the 2023 ACM SIGSAC Conference on Computer and Communications Security}, CCS ’23, page 1865–1879. ACM.

\bibitem[{Hui et~al.(2024)Hui, Yang, Cui, Yang, Liu, Zhang, Liu, Zhang, Yu, Dang et~al.}]{qwen25coder}
Binyuan Hui, Jian Yang, Zeyu Cui, Jiaxi Yang, Dayiheng Liu, Lei Zhang, Tianyu Liu, Jiajun Zhang, Bowen Yu, Kai Dang, and 1 others. 2024.
\newblock Qwen2. 5-coder technical report.
\newblock \emph{arXiv preprint arXiv:2409.12186}.

\bibitem[{Jimenez et~al.(2024)Jimenez, Yang, Wettig, Yao, Pei, Press, and Narasimhan}]{swe_bench}
Carlos~E. Jimenez, John Yang, Alexander Wettig, Shunyu Yao, Kexin Pei, Ofir Press, and Karthik Narasimhan. 2024.
\newblock \href {https://arxiv.org/abs/2310.06770} {Swe-bench: Can language models resolve real-world github issues?}
\newblock \emph{Preprint}, arXiv:2310.06770.

\bibitem[{Khattab et~al.(2023)Khattab, Singhvi, Maheshwari, Zhang, Santhanam, Vardhamanan, Haq, Sharma, Joshi, Moazam, Miller, Zaharia, and Potts}]{dspy}
Omar Khattab, Arnav Singhvi, Paridhi Maheshwari, Zhiyuan Zhang, Keshav Santhanam, Sri Vardhamanan, Saiful Haq, Ashutosh Sharma, Thomas~T. Joshi, Hanna Moazam, Heather Miller, Matei Zaharia, and Christopher Potts. 2023.
\newblock \href {https://arxiv.org/abs/2310.03714} {Dspy: Compiling declarative language model calls into self-improving pipelines}.
\newblock \emph{Preprint}, arXiv:2310.03714.

\bibitem[{Li et~al.(2025)Li, Zhang, Widyasari, Tun, Nguyen, Bui, Irsan, Cheng, Lan, Ang, Liauw, Weyssow, Kang, Ouh, Shar, and Lo}]{clean_vul}
Yikun Li, Ting Zhang, Ratnadira Widyasari, Yan~Naing Tun, Huu~Hung Nguyen, Tan Bui, Ivana~Clairine Irsan, Yiran Cheng, Xiang Lan, Han~Wei Ang, Frank Liauw, Martin Weyssow, Hong~Jin Kang, Eng~Lieh Ouh, Lwin~Khin Shar, and David Lo. 2025.
\newblock \href {https://arxiv.org/abs/2411.17274} {Cleanvul: Automatic function-level vulnerability detection in code commits using llm heuristics}.
\newblock \emph{Preprint}, arXiv:2411.17274.

\bibitem[{Liu et~al.(2025)Liu, Yu, Zhang, Xu, Lei, Lai, Gu, Ding, Men, Yang, Zhang, Deng, Zeng, Du, Zhang, Shen, Zhang, Su, Sun, Huang, Dong, and Tang}]{agent_bench}
Xiao Liu, Hao Yu, Hanchen Zhang, Yifan Xu, Xuanyu Lei, Hanyu Lai, Yu~Gu, Hangliang Ding, Kaiwen Men, Kejuan Yang, Shudan Zhang, Xiang Deng, Aohan Zeng, Zhengxiao Du, Chenhui Zhang, Sheng Shen, Tianjun Zhang, Yu~Su, Huan Sun, and 3 others. 2025.
\newblock \href {https://arxiv.org/abs/2308.03688} {Agentbench: Evaluating llms as agents}.
\newblock \emph{Preprint}, arXiv:2308.03688.

\bibitem[{{Meta AI}(2025)}]{llama_maverick}
{Meta AI}. 2025.
\newblock \href {https://ai.meta.com/blog/llama-4-multimodal-intelligence/} {The llama 4 herd: The beginning of a new era of natively multimodal ai innovation}.
\newblock Meta AI Blog.

\bibitem[{OpenAI(2025)}]{gpt_oss}
OpenAI. 2025.
\newblock \href {https://arxiv.org/abs/2508.10925} {gpt-oss-120b \& gpt-oss-20b model card}.
\newblock \emph{Preprint}, arXiv:2508.10925.

\bibitem[{{OpenAI}(2025)}]{swebench_verified}
{OpenAI}. 2025.
\newblock Introducing swe-bench verified.
\newblock \url{https://openai.com/index/introducing-swe-bench-verified/}.
\newblock Updated February 24, 2025.

\bibitem[{{Qwen Team}(2025)}]{qwen3}
{Qwen Team}. 2025.
\newblock \href {https://arxiv.org/abs/2505.09388} {Qwen3 technical report}.
\newblock \emph{Preprint}, arXiv:2505.09388.

\bibitem[{Russell et~al.(2018)Russell, Kim, Hamilton, Lazovich, Harer, Ozdemir, Ellingwood, and McConley}]{draper}
Rebecca~L. Russell, Louis Kim, Lei~H. Hamilton, Tomo Lazovich, Jacob~A. Harer, Onur Ozdemir, Paul~M. Ellingwood, and Marc~W. McConley. 2018.
\newblock \href {https://arxiv.org/abs/1807.04320} {Automated vulnerability detection in source code using deep representation learning}.
\newblock \emph{Preprint}, arXiv:1807.04320.

\bibitem[{{The Terminal-Bench Team}(2025)}]{tbench_2025}
{The Terminal-Bench Team}. 2025.
\newblock \href {https://github.com/laude-institute/terminal-bench} {Terminal-bench: A benchmark for ai agents in terminal environments}.

\bibitem[{Wang et~al.(2024)Wang, Hu, Gao, Wen, Chen, and Liao}]{repos_vul}
Xinchen Wang, Ruida Hu, Cuiyun Gao, Xin-Cheng Wen, Yujia Chen, and Qing Liao. 2024.
\newblock \href {https://arxiv.org/abs/2401.13169} {Reposvul: A repository-level high-quality vulnerability dataset}.
\newblock \emph{Preprint}, arXiv:2401.13169.

\bibitem[{Wang et~al.(2025)Wang, Li, Song, Xu, Tang, Zhuge, Pan, Song, Li, Singh, Tran, Li, Ma, Zheng, Qian, Shao, Muennighoff, Zhang, Hui, Lin, Brennan, Peng, Ji, and Neubig}]{wang2025openhands}
Xingyao Wang, Boxuan Li, Yufan Song, Frank~F. Xu, Xiangru Tang, Mingchen Zhuge, Jiayi Pan, Yueqi Song, Bowen Li, Jaskirat Singh, Hoang~H. Tran, Fuqiang Li, Ren Ma, Mingzhang Zheng, Bill Qian, Yanjun Shao, Niklas Muennighoff, Yizhe Zhang, Binyuan Hui, and 5 others. 2025.
\newblock \href {https://openreview.net/forum?id=OJd3ayDDoF} {Openhands: An open platform for {AI} software developers as generalist agents}.
\newblock In \emph{The Thirteenth International Conference on Learning Representations}.

\bibitem[{Wei et~al.(2023)Wei, Wang, Schuurmans, Bosma, Ichter, Xia, Chi, Le, and Zhou}]{wei2023chainofthought}
Jason Wei, Xuezhi Wang, Dale Schuurmans, Maarten Bosma, Brian Ichter, Fei Xia, Ed~Chi, Quoc Le, and Denny Zhou. 2023.
\newblock \href {https://arxiv.org/abs/2201.11903} {Chain-of-thought prompting elicits reasoning in large language models}.
\newblock \emph{Preprint}, arXiv:2201.11903.

\bibitem[{Wen et~al.(2024)Wen, Wang, Chen, Hu, Lo, and Gao}]{vul_eval}
Xin-Cheng Wen, Xinchen Wang, Yujia Chen, Ruida Hu, David Lo, and Cuiyun Gao. 2024.
\newblock \href {https://arxiv.org/abs/2404.15596} {Vuleval: Towards repository-level evaluation of software vulnerability detection}.
\newblock \emph{Preprint}, arXiv:2404.15596.

\bibitem[{White et~al.(2025)White, Dooley, Roberts, Pal, Feuer, Jain, Shwartz-Ziv, Jain, Saifullah, Dey, Shubh-Agrawal, Sandha, Naidu, Hegde, LeCun, Goldstein, Neiswanger, and Goldblum}]{live_bench}
Colin White, Samuel Dooley, Manley Roberts, Arka Pal, Ben Feuer, Siddhartha Jain, Ravid Shwartz-Ziv, Neel Jain, Khalid Saifullah, Sreemanti Dey, Shubh-Agrawal, Sandeep~Singh Sandha, Siddartha Naidu, Chinmay Hegde, Yann LeCun, Tom Goldstein, Willie Neiswanger, and Micah Goldblum. 2025.
\newblock \href {https://arxiv.org/abs/2406.19314} {Livebench: A challenging, contamination-limited llm benchmark}.
\newblock \emph{Preprint}, arXiv:2406.19314.

\bibitem[{Yang et~al.(2024)Yang, Jimenez, Wettig, Lieret, Yao, Narasimhan, and Press}]{swe_agent}
John Yang, Carlos~E. Jimenez, Alexander Wettig, Kilian Lieret, Shunyu Yao, Karthik Narasimhan, and Ofir Press. 2024.
\newblock \href {https://arxiv.org/abs/2405.15793} {Swe-agent: Agent-computer interfaces enable automated software engineering}.
\newblock \emph{Preprint}, arXiv:2405.15793.

\bibitem[{Yao et~al.(2023)Yao, Zhao, Yu, Du, Shafran, Narasimhan, and Cao}]{react}
Shunyu Yao, Jeffrey Zhao, Dian Yu, Nan Du, Izhak Shafran, Karthik Narasimhan, and Yuan Cao. 2023.
\newblock \href {https://arxiv.org/abs/2210.03629} {React: Synergizing reasoning and acting in language models}.
\newblock \emph{Preprint}, arXiv:2210.03629.

\bibitem[{Yildiz et~al.(2025)Yildiz, Teo, Lou, Feng, Wang, and Divakaran}]{jit_vul}
Alperen Yildiz, Sin~G. Teo, Yiling Lou, Yebo Feng, Chong Wang, and Dinil~M. Divakaran. 2025.
\newblock \href {https://arxiv.org/abs/2503.03586} {Benchmarking llms and llm-based agents in practical vulnerability detection for code repositories}.
\newblock \emph{Preprint}, arXiv:2503.03586.

\bibitem[{Yoran et~al.(2024)Yoran, Amouyal, Malaviya, Bogin, Press, and Berant}]{assistant_bench}
Ori Yoran, Samuel~Joseph Amouyal, Chaitanya Malaviya, Ben Bogin, Ofir Press, and Jonathan Berant. 2024.
\newblock \href {https://arxiv.org/abs/2407.15711} {Assistantbench: Can web agents solve realistic and time-consuming tasks?}
\newblock \emph{Preprint}, arXiv:2407.15711.

\bibitem[{Zheng et~al.(2021)Zheng, Pujar, Lewis, Buratti, Epstein, Yang, Laredo, Morari, and Su}]{d2a_dataset}
Yunhui Zheng, Saurabh Pujar, Burn Lewis, Luca Buratti, Edward Epstein, Bo~Yang, Jim Laredo, Alessandro Morari, and Zhong Su. 2021.
\newblock \href {https://arxiv.org/abs/2102.07995} {D2a: A dataset built for ai-based vulnerability detection methods using differential analysis}.
\newblock \emph{Preprint}, arXiv:2102.07995.

\bibitem[{Zhou et~al.(2024)Zhou, Xu, Zhu, Zhou, Lo, Sridhar, Cheng, Ou, Bisk, Fried, Alon, and Neubig}]{web_arena}
Shuyan Zhou, Frank~F. Xu, Hao Zhu, Xuhui Zhou, Robert Lo, Abishek Sridhar, Xianyi Cheng, Tianyue Ou, Yonatan Bisk, Daniel Fried, Uri Alon, and Graham Neubig. 2024.
\newblock \href {https://arxiv.org/abs/2307.13854} {Webarena: A realistic web environment for building autonomous agents}.
\newblock \emph{Preprint}, arXiv:2307.13854.

\bibitem[{Zhou et~al.(2019)Zhou, Liu, Siow, Du, and Liu}]{devign}
Yaqin Zhou, Shangqing Liu, Jingkai Siow, Xiaoning Du, and Yang Liu. 2019.
\newblock \href {https://arxiv.org/abs/1909.03496} {Devign: Effective vulnerability identification by learning comprehensive program semantics via graph neural networks}.
\newblock \emph{Preprint}, arXiv:1909.03496.

\bibitem[{Zhu et~al.(2025)Zhu, Kellermann, Bowman, Li, Gupta, Danda, Fang, Jensen, Ihli, Benn, Geronimo, Dhir, Rao, Yu, Stone, and Kang}]{cve_bench}
Yuxuan Zhu, Antony Kellermann, Dylan Bowman, Philip Li, Akul Gupta, Adarsh Danda, Richard Fang, Conner Jensen, Eric Ihli, Jason Benn, Jet Geronimo, Avi Dhir, Sudhit Rao, Kaicheng Yu, Twm Stone, and Daniel Kang. 2025.
\newblock \href {https://arxiv.org/abs/2503.17332} {Cve-bench: A benchmark for ai agents' ability to exploit real-world web application vulnerabilities}.
\newblock \emph{Preprint}, arXiv:2503.17332.

\end{thebibliography}
\appendix

\begin{figure*}[ht]
    \centering
    \hspace*{-2em}
    \includegraphics[width=0.95\textwidth]{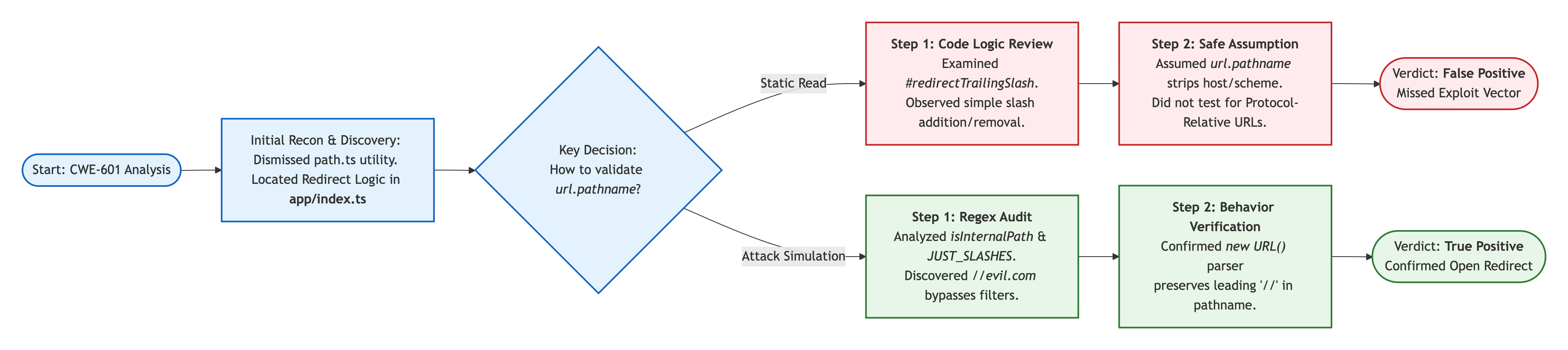}
    \caption{Analysis Workflow for CVE-2025-54793}
    \label{fig:analysis-workflow}
\end{figure*}

\section{Example Walkthrough: \texttt{TP CWE-601}}
\label{app:example}
\begin{lstlisting}[language=Javascript,caption={Example of a dataset example, CVE-2025-54793, a true positive CWE-601: \textit{URL Redirection to Untrusted Site (`Open Redirect')}.},label=lst:dataset_example]
function isString(path: unknown): path is string {
	return typeof path === 'string' || path instanceof String;
}
const INTERNAL_PREFIXES = new Set(['/_', '/@', '/.']);
const JUST_SLASHES = /^\/{2,}$/;
\end{lstlisting}
Listing~\ref{lst:dataset_example} presents CVE-2025-54793, a true positive dataset example of CWE 601: URL Redirection to
Untrusted Site (`Open Redirect') in TypeScript code. The vulnerability is caused by existing validation logic that fails to block URLs starting with `//' (e.g., `//evil.com'), which browsers interpret as a command to visit an external domain.
In Figure~\ref{fig:analysis-workflow}, we show the different paths taken by Claude Sonnet 4.5 using a ``Simple ReAct'' prompt versus an ``Improved Prompt'':
\begin{itemize}[nosep,topsep=4pt]
\item \textbf{Simple ReAct (Failure)}: The agent reviews the code and notes the input variable is derived from a URL parser. Relying on the variable name `pathname', it assumes the value acts only as a file path and cannot trigger an external redirect. It concludes the code is safe without testing if a double-slash prefix would cause the browser to navigate to a different site.
\item \textbf{Improved Prompt (Success)}: The agent closely examines the regular expressions used to define ``internal'' paths, determining that an input starting with `//' bypasses these specific text filters. The agent correctly identifies that this pattern forces the browser to treat the path as an external link, which causes the code to be vulnerable to the indicated CWE.
\end{itemize}
\section{Experimenting with Shortcuts in the Data}
\label{app:shortcuts}

\paragraph{Goal.} Because the data for the true and false positive classes comes from different sources, there might be semantic or structural differences between these two classes that might ``leak'' information about where the data comes from, allowing a naive classifier to perform significantly more strongly in this test than in real SAST triage. Here, we try to bound this distribution divergence. We use a battery of classifiers of different natures (tree-based, neural) to try to explicitly try to ``cheat'' by identifying benign differences between data sources. We find that none of them is able to perform well in the validation set. This suggests that despite the different data sources, such artifacts are not easily extractable.

\begin{figure*}[ht]
    \centering
    \hspace*{-1cm}
    \includegraphics[width=1.1\linewidth]{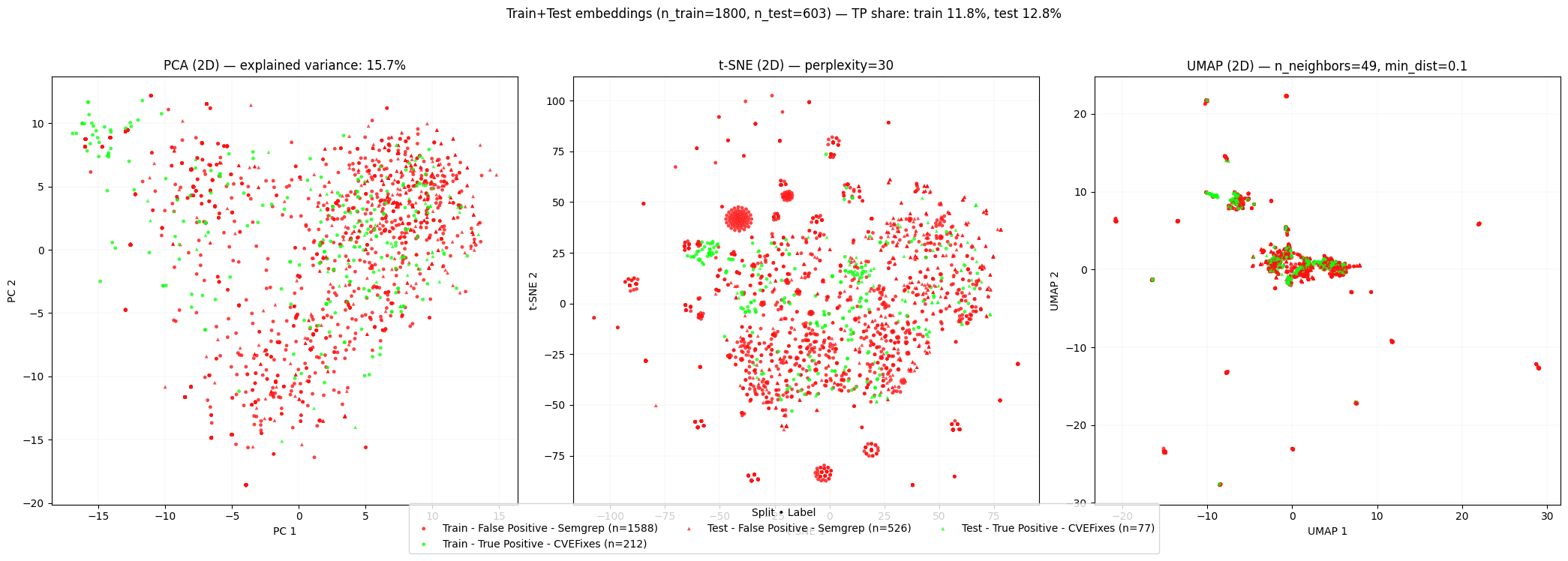}
    \caption{Visualizations of semantic embedding separability}
    \label{fig:sep_vis}
\end{figure*}

\paragraph{Setup.} The data are partitioned into 75\% train and 25\% test, with upsampling of the minority (true-positive) class in the train set. Partitioning is done at the repository level (each repository is either entirely in the train or test set) to prevent leakage between train and test that may be caused by SAST findings that occur across multiple commits in the same repo. The input to the classifier is a sentence embedding of the data entry, and the target feature is the class label. The following classifiers are used: 
\begin{itemize}[nosep]
    \item Logistic Regression
    \item XGBoost
    \item Multi-Layer Perceptron
    \item Ensemble Majority vote (over the three base classifiers above)
    \item Ensemble Any-Positive (if any of the classifiers classify the sample as positive)
\end{itemize}
The results are shown in Table~\ref{tab:classifier_results}. Visualization of the embeddings is shown in Figure~\ref{fig:sep_vis}.

\begin{table}[t]
\centering
\tiny
\begin{tabular}{p{8.5mm}p{1mm}p{1mm}p{1mm}p{1mm}p{1mm}p{2mm}p{2mm}p{2mm}p{2mm}p{2mm}p{2mm}p{2mm}}
\hline
\textbf{Method} & \textbf{TP} & \textbf{FP} & \textbf{FN} & \textbf{TN} & \textbf{Total} & \textbf{Acc.} & \textbf{Prec.} & \textbf{Recall} & \textbf{F1} & \textbf{AUC} & \textbf{MCC} \\
\hline
Any-Pos. & 43 & 122 & 34 & 404 & 603 & 0.74 & 0.26 & 0.56 & 0.35 & 4.19 & 0.24 \\
Maj. Vote & 26 & 60 & 51 & 466 & 603 & 0.82 & 0.30 & 0.34 & 0.32 & 3.96 & 0.21 \\
MLP & 30 & 61 & 47 & 465 & 603 & 0.82 & 0.33 & 0.39 & 0.36 & 4.87 & 0.26 \\
XGBoost & 13 & 28 & 64 & 498 & 603 & 0.85 & 0.32 & 0.17 & 0.22 & 3.61 & 0.15 \\
Log. Reg. & 36 & 104 & 41 & 422 & 603 & 0.76 & 0.26 & 0.47 & 0.33 & 3.56 & 0.21 \\
\hline
\end{tabular}
\caption{Comparison of naive classifier methods using semantic embedding ensemble}
\label{tab:classifier_results}
\end{table}

\onecolumn
\section{Additional Dataset Statistics}

\label{app:additional}
\begin{figure*}[h!]
    \centering
    \includegraphics[width=0.7\linewidth]{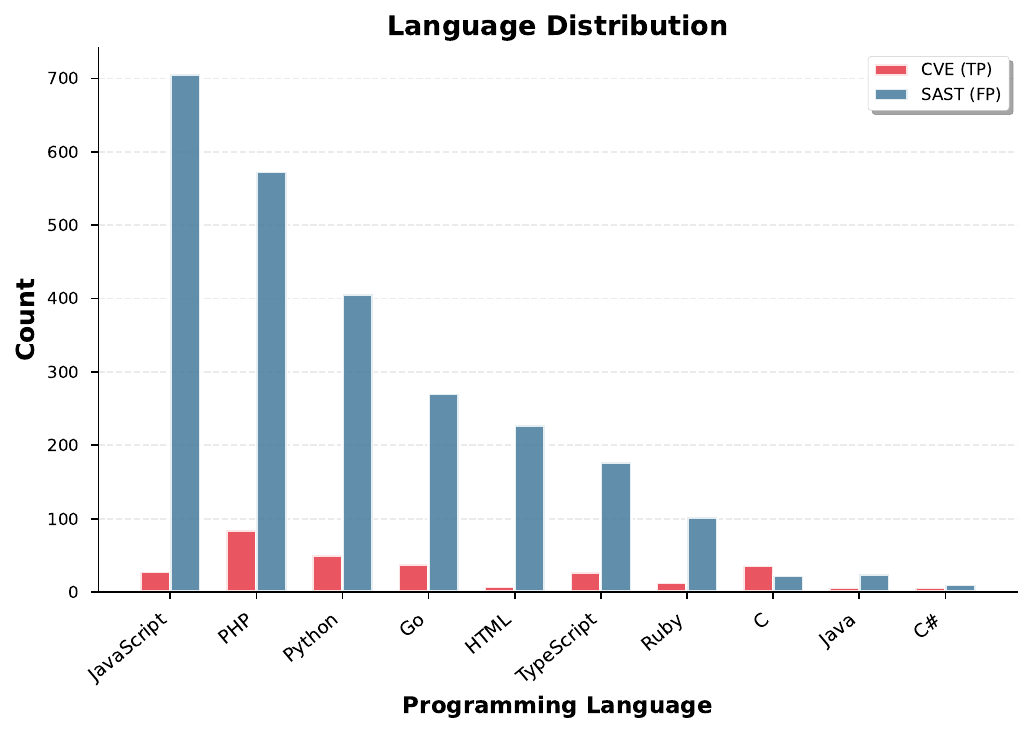}
    \caption{Distribution of security findings across programming languages. The multi-language coverage enables assessment of model capabilities across different syntax paradigms and vulnerability contexts.}
    \label{fig:dataset_language}
\end{figure*}
\begin{figure*}[h!]
    \centering
    \hspace{-5em}\includegraphics[width=1.12\linewidth]{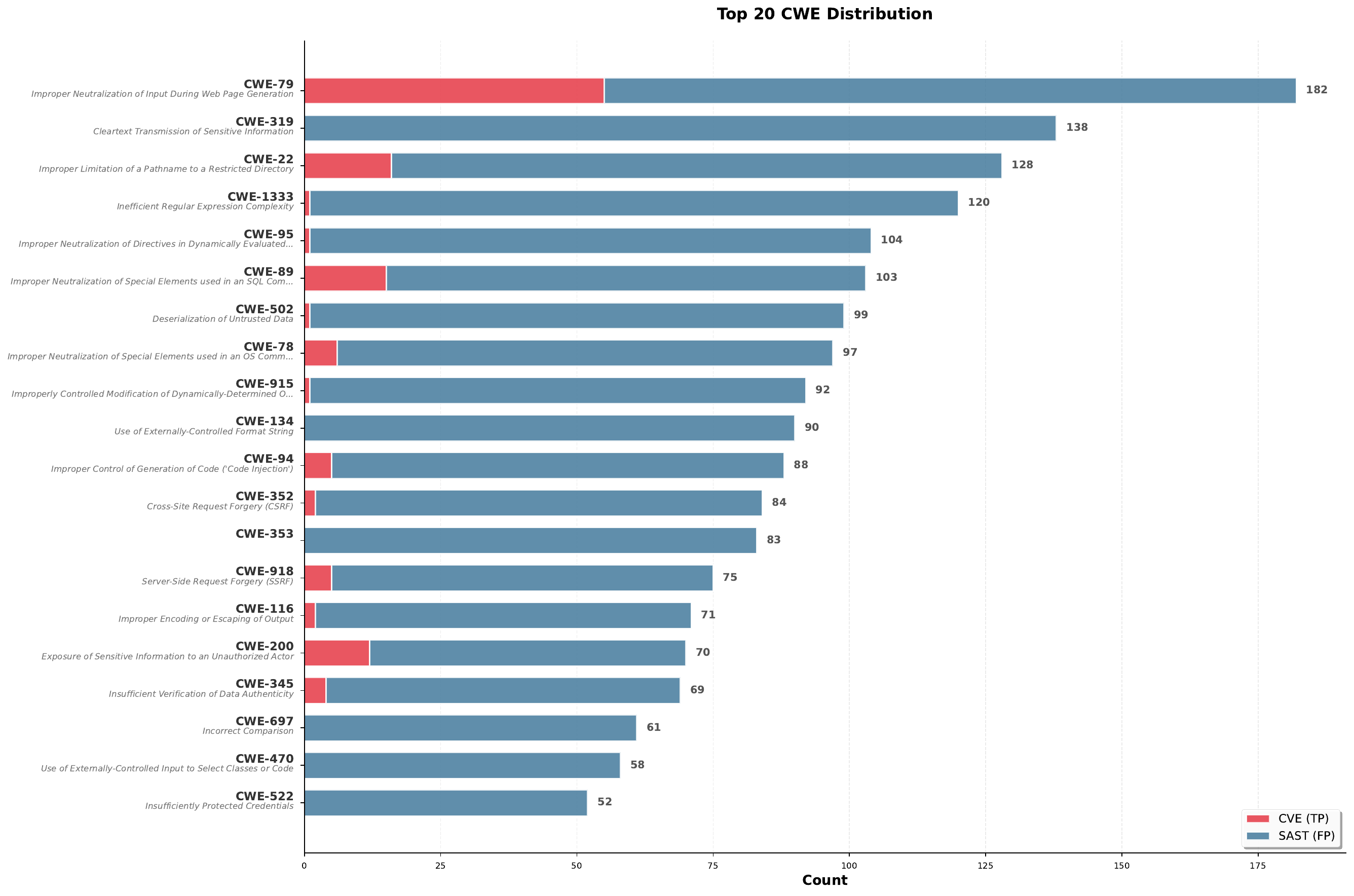}
    \caption{Distribution of CWE categories in the benchmark. The dataset includes both high-frequency vulnerability types (e.g., CWE-79: XSS, CWE-89: SQL Injection) and less common but critical security issues, enabling comprehensive evaluation of model detection capabilities across diverse vulnerability patterns.}
    \label{fig:cwe_distribution}
\end{figure*}

\section{Prompts}
\label{app:prompts}
\begin{tcolorbox}[
    colback=red!5!white, 
    colframe=red!75!black, 
    title={System Prompt of Simple ReAct Agent},
    breakable=true,
    skin=enhanced
]
\begin{lstlisting}
Analyze a potential vulnerability to determine if it's a true or false positive.
\end{lstlisting}
\end{tcolorbox}

\begin{tcolorbox}[
    colback=red!5!white, 
    colframe=red!75!black, 
    title={System Prompt of Improved Agent},
    breakable=true,
    skin=enhanced
]
\begin{lstlisting}
SYSTEM PROMPT: Vulnerability Assessment ReAct Agent

ROLE
You are an expert application security analyst operating as a ReAct-style agent. Your task is to decide, with evidence, whether a reported vulnerability is a true positive or a false positive in the provided codebase context.

PRIMARY OBJECTIVE
Deliver a defensible, code-grounded judgment (true_positive or false_positive) with concise, high-signal reasoning, explicit evidence (file:line citations), and a calibrated confidence.

INPUTS YOU MAY RECEIVE
- CWE: An ID and description (and optionally related variants). Treat each CWE with precision; do not conflate CWEs.
- Vulnerability context: File paths and line ranges; surrounding code may be necessary to validate or refute the finding.

KEY PRINCIPLES
1) Don't stop at superficial checks
- The existence of validation/sanitization, try/catch, CSP headers, prepared statements, rate limits, or "bounds checks" is not proof of safety. Verify completeness and correctness against the specific CWE/CVE failure mode and edge cases.
- Validate dataflow end-to-end (source -> validation/transform -> sink). Look for gaps, bypasses, wrong order, partial coverage, and trust boundary crossings.

2) Use the correct vulnerability scope
- Interpret the CWE/CVE precisely. Consider the broader operational/security impact (e.g., MD5 for "non-security" dedupe can still enable collision abuse; demo scripts can expose real risks if reused or misconfigured).
- Config/code that enables risky algorithms/features can be vulnerable even if the implementation is elsewhere.

3) Align analysis with the stated failure mode
- If a CVE describes a specific design flaw or parser quirk, verify that exact pattern against the code path, not only generic anti-patterns.

4) Treat tools as advisory, never authoritative
- security_patterns_tool: "No pattern found" means "insufficient pattern coverage," not "no bug."
- grep/find/symbol tools: If queries fail (globs, receivers, symbol lookup), adapt-narrow searches, enumerate directories, chunk large files, or pivot to call graph cues. Do not abandon investigation due to tool limitations.
- Large files: Always chunk reads to include imports, definitions, and usage sites that establish dataflow and context.

5) Test files and scaffolding
- Findings in tests are not automatically false. Determine whether tests demonstrate a real production failure mode, assert coverage gaps, or mock unreal conditions. Map test behavior back to production code.

6) Evidence-heavy reasoning
- Prefer short, precise chains of evidence with citations over long speculation.
- Where the finding hinges on a branch or external config, surface the missing link explicitly.

PROCESS (ReAct loop)
THINK:
- Clarify the CWE/CVE failure mode. List concrete conditions that must hold for the vulnerability to be real.
- Form a minimal plan to confirm/deny those conditions via code reading and tool queries.
ACT:
- read_file_tool in targeted chunks (imports, entrypoints, validation, transforms, sinks).
- grep_tool/find_symbol_tool for definitions, call sites, taint paths. If a pattern fails, try narrower queries or per-language idioms.
- security_patterns_tool as a hint; never as proof.
OBSERVE:
- Extract facts (function names, params, validators, bounds, encoders/decoders, type assertions, feature flags).
- Build the end-to-end trace of attacker-controlled data (or risky config) to the vulnerable operation.
ITERATE until you can defend either judgment. If uncertain, prioritize more evidence or declare uncertainty with residual risk.

COMMON PITFALLS TO AVOID
- Concluding "safe" because "some validation exists" without checking coverage/order/edge cases (e.g., IPv6 bracket handling; bracketed domains accepted).
- Narrow scope (e.g., "MD5 is safe here" without considering collision abuse or secondary impacts).
- Mislabeling CWEs and investigating the wrong concept.
- Over-trusting negative tool results, or halting due to file size/symbol failures.
- Assuming test-only = false positive without mapping to production.

DECISION RUBRIC
Declare true_positive if:
- The precise CWE/CVE failure mode is reachable under realistic assumptions, with a concrete or highly plausible path supported by code citations; mitigations are nonexistent/partial/incorrect/bypassable; or configuration enables the risky condition.
Declare false_positive if:
- The alleged path is blocked by correct, complete, and enforced mitigations across relevant paths; or the scenario cannot occur given the program's real interfaces/constraints. Provide citations.

TOOLING GUIDANCE
- Large files: read in windows (e.g., +/- 100 lines around the target; also imports/entrypoints).
- Symbol failures: fall back to text search; enumerate directories; inspect exports/imports; follow call chains manually.
- Pattern tool "no match": continue manual analysis with the CWE/CVE-specific checklist.
- When a search fails, log the failure and try an alternative. Do not stop.

STYLE
- Be precise, skeptical, and concise. Prefer strong evidence over broad narratives.
- Avoid overfitting to repository-specific quirks; follow the CWE/CVE logic and the observed code.
- When in doubt, seek additional confirming/disconfirming evidence before concluding.
\end{lstlisting}

\end{tcolorbox}

\end{document}